\documentclass[a4paper,twoside]{article}

\usepackage{tikz}
\usepackage{adjustbox}
\definecolor{blue_shap}{RGB}{0,139,251}

\usepackage[hidelinks]{hyperref}
\usepackage{microtype}

\usepackage{pgfplots}
\usepgfplotslibrary{groupplots, colorbrewer}
\pgfplotsset{compat=1.3, cycle list/Dark2-5}

\usepackage[misc]{ifsym}

\usepackage{epsfig}
\usepackage{subcaption}
\usepackage{calc}
\usepackage{amssymb}
\usepackage{amstext}
\usepackage{amsmath}
\usepackage{amsthm}
\usepackage{multicol}
\usepackage{pslatex}
\usepackage{apalike}
\usepackage[bottom]{footmisc}
\usepackage{SCITEPRESS}     

\newcommand{\numjavalibs}{13 }
\newcommand{\numnativelibs}{10 }
\newcommand{\numanalyzedclasses}{86 }
\newcommand{\numstatements}{333 }
\newcommand{\numsamplesdetailed}{$603\,937$ }
\newcommand{\nummalicious}{$301\,898$ }
\newcommand{\numbenign}{$302\,039$ }

\begin{document}

\title{A Longitudinal Study of Cryptographic API: a Decade of Android Malware}

\author{\authorname{Adam Janovsky\sup{1}(\Letter), Davide Maiorca\sup{2}, Dominik Macko\sup{1}, Vashek Matyas\sup{1} and Giorgio Giacinto\sup{2}}
\affiliation{\sup{1}Masaryk University, Czech Republic}
\affiliation{\sup{2}University of Cagliari, Italy}
\Letter~\email{adamjanovsky@mail.muni.cz}
}


\abstract{Cryptography has been extensively used in Android applications to guarantee secure communications, conceal critical data from reverse engineering, or ensure mobile users' privacy. Various system-based and third-party libraries for Android provide cryptographic functionalities, and previous works mainly explored the misuse of cryptographic API in benign applications. However, the role of cryptographic API has not yet been explored in Android malware. This paper performs a comprehensive, longitudinal analysis of cryptographic API in Android malware. In particular, we analyzed \numsamplesdetailed Android applications (half of them malicious, half benign) released between $2012$ and $2020$, gathering more than 1 million cryptographic API expressions. Our results reveal intriguing trends and insights on how and why cryptography is employed in Android malware. For instance, we point out the widespread use of weak hash functions and the late transition from insecure DES to AES. Additionally, we show that cryptography-related characteristics can help to improve the performance of learning-based systems in detecting malicious applications.}

\onecolumn \maketitle \normalsize \setcounter{footnote}{0} \vfill

\section{\uppercase{Introduction}}
\label{sec:intro}

The Android operating system has spread worldwide during the last decade, reaching almost $3$ billion users in $2021$~\cite{statista}. At the same time, security threats against Android have multiplied, as recent reports showed~\cite{mcfee19}. 

Most Android applications employ cryptographic primitives to conceal critical information and securely carry out communication with internal components, applications, and web services. At the same time, it is natural to imagine that malware authors may leverage cryptography in a plethora of artful ways to serve their malevolent objectives. For instance, cryptography equips attackers with the ability to fingerprint the parameters of an infected device, encrypt users' media files, establish a secure connection with a command-and-control server, or manage ransom payments carried out by victims infected by, e.g., ransomware. 

Previous research conveyed a significant effort in analyzing cryptography in benign applications. The focus was mainly related to the misuse of cryptographic application programming interface (API) in benign Android applications, i.e., on finding and eliminating vulnerabilities in the employed crypto-routines that may allow attackers to obtain sensitive information~\cite{egeleEmpiricalStudyCryptographic2013,muslukhovSourceAttributionCryptographic2018,chatzikonstantinouEvaluationCryptographyUsage2016,shuaiModellingAnalysisAutodetection2014}.

To the best of our knowledge, however, no study explored how cryptography is currently employed in \emph{malicious} applications. In this paper, we answer two important research questions related to cryptography and Android malware: 
\begin{enumerate}
    \item \textbf{RQ.1:} Are there significant differences in how cryptography is employed in benign and malicious applications?
    \item \textbf{RQ.2:} Can information about cryptography improve Android malware detection?
\end{enumerate}

We believe that answering these questions will shed more light on the mechanisms of Android malware, providing new insights for its analysis, characterization, and detection. In this paper, we propose two main contributions. First, we deliver a comprehensive comparison of how cryptography is employed in \numsamplesdetailed malicious and benign applications released in the last decade. Such a comparison is carried out with an open-source\footnote{The code is acessible from \url{github.com/adamjanovsky/AndroidMalwareCrypto}.}, scalable approach that inspects (among others) the usage of hash functions, symmetric and public-key encryption, PRNGs, etc. In total, we inspect over $10^6$ of cryptographic API expressions. 

Second, we show that cryptographic features can be used to augment the performances of state-of-the-art malware detectors, demonstrating their discriminant power in distinguishing malicious and benign applications. We also employ techniques inherited from the interpretation of learning models to point out possible connections between cryptographic API and malicious actions.

The attained results show many intriguing and surprising trends. 
For example, in contrast to benign applications, malware authors do not typically resort to strong cryptography to perform their actions. We show that malware often favors the use of cryptographically defeated primitives, e.g., weak hash functions MD5~\cite{wang2005break} or SHA-1~\cite{stevens2017first}, or symmetric encryption scheme DES~\cite{biham1991differential}. These insights can also be especially useful to learning-based models, which can leverage these cryptographic trends to improve the detection rate of malware. 
We believe that the results presented in this work can constitute a seminal step to foster additional research on the relationship between cryptography and Android malware.  

The paper is organized as follows: In Section~\ref{sec:methodology}, we describe the methodology of our analysis. In Section~\ref{sec:expresults}, we answer the first research question. Section~\ref{sec:expresults:subsec:rq2}  discusses cryptographic API in relation to malware detection. The limitations of our study are discussed in Section~\ref{sec:discussion}. Section~\ref{sec:relwork} describes the related work and Section~\ref{sec:conclusions} closes the paper with the conclusions.

We also deliver an extended technical report for this paper on arXiv\footnote{\url{https://arxiv.org/abs/2205.05573}.}.

\section{\uppercase{Technical Background}}
\label{sec:background}

In this section, we provide technical elements that allow the reader to understand the remaining content of this paper. We start by describing the main characteristics of Android applications. Then, we provide an overview of the techniques with which Android applications can be analyzed. Finally, we describe the prominent functionalities of the cryptographic API that can be employed in Android applications.

\subsection{Background on Android}
Android applications can be represented as zipped \texttt{.apk} (Android application package - APK) archives that feature: \emph{(i)} The \texttt{AndroidManifest.xml} file, which provides the application package name, the name of the app basic components, and the permissions that are required for specific operations; \emph{(ii)} One or more \texttt{classes.dex} files, which represent the  application executable(s), and which contain all the implemented classes and methods executed by the app. This file can be disassembled to a simplified format called \texttt{smali}; \emph{(iii)} Various \texttt{.xml} files that characterize the application layout; \emph{(iv)} External resources that include images and native libraries.

Although Android applications are typically written in \texttt{Java}, they are compiled to an intermediate bytecode format called \texttt{Dalvik}, whose instructions are contained in the \texttt{classes.dex} file. This file is parsed at install time and converted to a native ARM code that is executed by the Android RunTime (ART). In this way, it is possible to speed up the execution in comparison to the previous runtime (\texttt{dalvikvm}, available till Android $4.4$), in which applications were executed with a just-in-time approach (during installation, the \texttt{classes.dex} file was only slightly optimized, but not converted to native code).

\subsection{Analysis Techniques}
\label{sec:background:subsec:analysis}

Android applications can be analyzed either statically or dynamically. Static analysis can be performed either by disassembling the Dalvik bytecode's instructions or by decompiling the executable to its Java source. Typical analysis techniques involve, among others, program slicing, data-flow and taint analysis, and the extraction of the application call graphs. Dynamic analysis can be performed by tracing the execution of the instructions, as well as the changes that are presented in memory and during the execution of the applications.

Both approaches feature their limitations. Static analysis is particularly prone to be evaded by obfuscation techniques, such as renaming user-implemented functions, modifying the call graph, using reflection or encryption API~\cite{maiorca15-cose,hoffmann16-sac}.
Dynamic analysis can be especially challenging to carry out due to the so-called \emph{path-explosion} problem, where the application should be stimulated to take different execution branches. This operation is particularly complex in Android apps, as there are numerous ways to interact with them. As applications are typically executed in emulated environments, malicious applications can especially check whether the application is being debugged or not. Finally, dynamic analysis can be resource- and time-consuming.

The choice of the right technique explicitly depends on the analysis goals. For large-scale analyses, static analysis is typically recommended, as it is much faster to carry out, and its scalability outweighs its limitations over a large number of samples.

\subsection{Cryptography in Android}

Android developers typically have several means to implement cryptographic functionality for their applications: \emph{(i)} Using Java Cryptographic Architecture (JCA) via Android API; \emph{(ii)} Using third-party Java cryptographic libraries; \emph{(iii)} Using third-party native cryptographic libraries; \emph{(iv)} Designing and/or employing own cryptographic functions. Note that the last method is widely discouraged in the cryptographic community and is unfeasible to be reliably employed, as it has no well-grounded fingerprint. We also stress that, in most cases, developers do not need to develop their cryptographic functions, but they prefer to employ readily available tools. For these reasons, we will not further discuss the \emph{(iv)} case in the rest of the section.  

In Android API, the cryptographic functionality is delivered via JCA. The JCA provides a stable set of classes and functions that can be called from two main packages, \texttt{javax.crypto} and \texttt{java.security}~\cite{javatmplatformJavaCryptographyArchitecture2017,googleAndroidCryptographyAPI2020}. These packages contain more than 100 classes covering the majority of cryptographic primitives and protocols, such as hash functions, symmetric encryption schemes, digital signature algorithms, and so forth. Although the Android documentation explicitly recommends the use of specific primitives\footnote{AES-256 in CBC or GCM mode, SHA-2 for hash functions, SHA-2 HMAC for MACs and SHA-2 ECDSA for signatures as of early 2020~\cite{googleAndroidCryptographyAPI2020}.}, many insecure cryptographic primitives (such as the MD5 hashing function or the symmetric cipher DES) can be chosen. Apart from providing the API, the JCA introduces an abstraction layer of the so-called Cryptographic Service Providers. Such providers register themselves at the JCA and are responsible for implementing any subset of the API. Different version of Android prefer different providers, we for instance mention \texttt{BouncyCastle}~\cite{legionofthebouncycastleinc.LegionBouncyCastle} and \texttt{Conscrypt}~\cite{conscrypt} that are among the most popular ones. While these providers differ in their internals, they must all comply with the API and expose identical function names and argument ranges. 

Outside the Android API, practically any cryptographic functionality can be supplied by some third-party library. Still, there is no curated list of neither Java nor native cryptographic libraries for the Android platform to our best knowledge. During our study, we witnessed libraries that have specific focuses, such as providing functionality only for AES encryption. On the contrary, there are also libraries that expose various cryptographic primitives and protocols, such as the OpenSSL library. These full-fledged libraries can supply similar functionality to the Android API.

\section{\uppercase{Methodology}}
\label{sec:methodology}

This section describes the methodology we employed to extract and analyze the cryptographic API embedded in Android applications. We start by formalizing the problem, properly defining its domain and various constraints. We then show how we implemented this formalism in practice by discussing our developed analysis framework. 
Our findings are based on the static analysis of the Java source code obtained by the decompilation of the Android executables.

\subsection{Problem Formalization}
\label{sec:methodology:subsec:formalization}

We organize the problem formalization in two parts: part one treats the definition of the crypto-routines of interest for our analysis, part two describes the process of locating those routines in the application source code. 

\textbf{I. Definition of Crypto-Routines.}
Given a set of Android applications, we denote the set of all possible functions  $\mathbb{F}$ contained in their source code as: 

\begin{equation*}
 \mathbb{F} = \mathbb{U} \cup \mathbb{S} \cup \mathbb{T} = \mathbb{C} \cup \mathbb{C}^{c},
\end{equation*}
Where $\mathbb{U}$ represents the set of functions defined by the user, $\mathbb{S}$  is the set of system-related functions contained in the Android SDK, and $\mathbb{T}$ is the set of functions belonging to third-party libraries. Given a set of known crypto-related functions $\mathbb{C}$, our goal is to study the intersection of $\mathbb C$ and $\mathbb S$, denoted as $\mathbb F_{cs}$. In other words, $\mathbb{F}_{cs}$ is the set of cryptography-related functions that are defined in the system package (in Android represented by JCA functions). Notably, in this analysis, we discard custom cryptographic functions that users or third parties may implement. Automatic detection of such functions would be a complex task in the context of a large-scale analysis, which may lead to false positives (or negatives) without further manual inspection. In our study, we solely aim to answer what functions from $\mathbb F_{cs}$ the malware authors favor.

From the cryptographical perspective, the functions contained in $\mathbb{F}_{cs}$ can be divided into the following categories: 
\textit{(i) Hash functions}. Cryptographic hash functions such as MD5, SHA-1, or SHA-2; 
\textit{(ii) Symmetric encryption}. Symmetric cipher primitives such as AES, DES, or RC4;  
\textit{(iii) Public-key encryption}. Asymmetric primitives, in Android represented by the RSA cryptosystem; 
\textit{(iv) Digital signature algorithms}. Primitives that empower digital signatures, e.g., ECDSA;
\textit{(v) MAC algorithms}. Primitives that construct Message Authentication Codes, also called MACs;
\textit{(vi) PRNG}. Functions to run pseudo-random number generators (PRNG);
\textit{(vii) Key agreement protocols}. Algorithms for key exchange, in JCA represented by Diffie-Hellman protocol;
\textit{(viii) Others}. Functions that do not fall into any of the previous categories.

\textbf{II. Locating Cryptographic API.}
All functions in $\mathbb{F}_{cs}$ are available through two Java packages in Android API: \texttt{javax.crypto} and \texttt{java.security}. Our research aims to reveal \emph{which cryptographic functions have been chosen and directly employed by the authors}. Notably, Android applications typically contain third-party packages that invoke crypto functions. We aim to exclude those packages from our analysis as the application authors did not contribute to them.

Thus, for each Android sample, we are interested in extracting the cryptographic API $\mathbb{F}_{a} \subseteq \mathbb{F}_{cs}$ that is invoked from user-defined functions $\mathbb U$. To obtain the functions belonging to $\mathbb{F}_{a}$, we perform two steps: \emph{(i)} We automatically detect the classes that belong to third-party or system libraries, and we exclude them from the set of classes that should be explored. By doing so, we establish the list of \emph{user-implemented functions} $\mathbb U$; \emph{(ii)} We extract all references to crypto-related functions $\mathbb F_{cs}$ that are invoked directly from $\mathbb U$. 

The first step is motivated by the discovery~\cite{wangWuKongScalableAccurate2015} that more than 60\% of Android APK\footnote{Android Application Package, an archive that encapsulates the whole Android application.}
code (on average) originates from third-party packages. To study user-authored code, it is therefore critical to differentiate, with reasonable certainty, whether a class belongs to a third-party library or not. This task can be extremely challenging and was extensively studied, e.g., by~\cite{wangWuKongScalableAccurate2015,maLibRadarFastAccurate2016,libscout}. It does not suffice to merely search for the \texttt{import} clauses in the decompiled source code since the non-system packages could be renamed. This scenario is especially frequent in malicious applications, as the authors aim to defend against any forensics. Inspired by the systematic review of third-party package detectors~\cite{zhan_automated_2020}, we opted to tackle this with \texttt{LibRadar}, a popular third-party libraries detection tool that utilizes clustering and complex signatures to recognize such libraries~\cite{maLibRadarFastAccurate2016}. In this review,  \texttt{LibRadar} achieves the highest precision and second-highest recall while it takes approx. 5 seconds to evaluate an APK on average. The runner-up requires over 80 seconds per APK, which would be unsuitable for large-scale analysis. \texttt{LibRadar} was trained on a large dataset of Android applications and can reliably fingerprint more than $29\,000$ third-party libraries, not relying on package names. Consequently, \texttt{LibRadar} can identify obfuscated packages. Using \texttt{LibRadar}\footnote{Since \texttt{LibRadar} requires large Redis database to run (preventing parallelization), we actually leveraged its lightweight version \texttt{LiteRadar}. Prior to doing so, we compared the output of both tools on a small subset to find out that this decision has a negligible effect on the number of detected libraries.}, we filter the identified third-party packages of an APK from subsequent cryptographic API analysis.

\subsection{Crypto API Extraction Pipeline}
\label{sec:methodology:subsec:pipeline}
Our system generates a comprehensive report of the embedded cryptographic API, given an application dataset.
As an input, configuration files for the to-be-conducted experiment are taken. Apart from other choices, the files contain a list of APKs that can be loaded from a disk or downloaded from the Internet. 

The APKs are then processed in parallel, and each sample traverses the following pipeline:
\begin{enumerate}
    \item \textbf{Pre-processor}. 
    This module decompiles the APKs to obtain their Java source code. Then, the third-party packages of the APKs are identified, and the whole Java source code of the APKs is extracted. 
    \item \textbf{Crypto-extractor}. This module extracts and analyzes the cryptographic function call sites in the application source code. Their filtering is achieved by matching pre-defined regular expressions. Additionally, the crypto-extractor also detects both Java and native third-party cryptographic libraries. 
    \item \textbf{Evaluator}. This module stores, organizes, and aggregates the information retrieved by the analyzed APKs to a JSON record.
\end{enumerate} 
The evaluator outputs a report of the cryptographic usage for each APK. We designed the system in a modular fashion so that one can alter its inner workings to extract further valuable insights from the APKs. 

\subsection{Dataset}
\label{sec:methodology:subsec:dataset}
To gain an all-around view of the cryptographic API landscape in Android applications, we leverage the Androzoo dataset~\cite{allixAndroZooCollectingMillions2016}. Currently, Androzoo is the largest available dataset of Android applications, containing more than 15 million of APKs. We sampled \numbenign benign applications and \nummalicious malicious applications from Androzoo released in the years 2012-2020. 
We strived for uniform distribution of samples in the studied timeline. Yet, for years 2018, 2020 we could only collect a limited number of malicious samples -- $19\,305$ and $10\,039$, respectively. To speed up the computation, we only gathered APKs smaller than $20\,\textrm{MB}$ (approximately 89\% of malicious APKs in the Androzoo fulfill this criterion). 

To accurately discriminate malicious files, we consider an APK as malicious if it was flagged malicious by at least five antivirus scanners from the VirusTotal service\footnote{\url{virustotal.com}. The number of VirusTotal positive flags is already contained in the Androzoo dataset.}, which should reliably eliminate benign files deemed malicious, as reported by Salem~\cite{salem2020towards}. 
Our samples are predominantly originating from 3 distinct sources: Google Play ($60\%$), Anzhi ($19\%$), and Appchina ($13\%$). Note that the samples were deduplicated on a per-market basis~\cite{allixAndroZooCollectingMillions2016} to avoid over-counting. 

\subsection{Cryptography and Machine Learning}
\label{sec:methodology:subsec:motivation}
Statistics about cryptographic usage are undoubtedly helpful in pointing out differences between benign and malicious applications. Another intriguing question to explore is \emph{whether such statistics can be useful to recognize malicious samples from benign ones effectively}. To answer this question, we propose three methods that employ machine learning techniques, described in the following.

\subsubsection{Cryptographic Learning Model} The first technique consists of defining a learning-based system whose structure is inspired by other popular detection systems~\cite{rieck14-drebin,chen16-asiaccs,maiorca_r-packdroid_2017}. In particular, the proposed system performs the following steps: \emph{(i)} it takes as an input an Android application and extracts its cryptographic API usage with the pipeline described in Section~\ref{sec:methodology:subsec:pipeline}; \emph{(ii)} it encodes this statistics into a vector of \emph{features}; \emph{(iii)} it trains a machine-learning classifier to predict a benign/malicious label.  

The feature vector includes features that can be categorized into three sets:
\begin{itemize}
    \item \textbf{Set A}: flags indicating the use of third-party cryptographic libraries (both Java and native). 
    \item \textbf{Set B}: frequencies of specific cryptographic API constructors and imports of crypto-related classes, e.g., number of DES constructors in a sample. 
    \item \textbf{Set C}: aggregated statistics of call sites and imports related to categories of cryptographic primitives: hash functions, symmetric encryption schemes, and so forth. For example: how many distinct hash functions a sample uses. 
\end{itemize}

By joining these sets, we obtain 300 potentially informative features. These features are further filtered with a feature selection algorithm. The dataset with candidate features is split in a 9:1 ratio into training/test sets. Then, we apply two feature selection methods to drop uninformative features. First, we examine all possible pairs of features. If a pair exhibits Pearson's correlation coefficient higher than 0.95, we drop a random feature of such a pair. Second, we remove the features that are deemed uninformative by Boruta~\cite{borutaFeatureSelection}. Boruta is a supervised algorithm that iteratively replicates features, randomly permutates their values, trains a random forest, and removes redundant features based on the z-score. The feature selection process yields 189 features that are used for learning. 

To choose the best family of models for discriminating between malicious and benign samples on our dataset, we experimented with naive Bayes, logistic regression, support vector machines with linear kernel, random forest, gradient boosted decision trees, and multilayer perceptron. We tuned the classifiers' hyperparameters using 10-fold cross-validation on the training dataset, optimizing for the F1 score. Subsequent evaluation yielded a random forest (which works as a majority-voting ensemble of decision trees trained on different subsets of the data) as the best-performing classifier (w.r.t. F1 score). We do not report the entire analysis here for brevity, and we stick to the random forest in the rest of the paper.

\subsubsection{Explaining the Learning Model}

To further advance the understanding of cryptographic API in Android detection, we explain the predictions of the just described cryptographic classifier. To do so, we follow the techniques of explainable machine learning, a field that attempts to understand why a classifier makes certain decisions. We study both \emph{global} and \emph{local} importances of the features extracted by the classifier, in a similar fashion to what has been proposed in previous work~\cite{melis18-eusipco,melis22-ijmlc}. A global analysis evaluates the impact of the features averaged over the whole dataset, while local analysis evaluates the impact of the features on specific samples.

To interpret the classifier's predictions, we used Shapley additive explanations (SHAP)~\cite{unifiedInterpretationApproach} that are successfully used outside of computer science.
SHAP can consistently explain both local predictions and global feature importance by measuring each feature's contribution to the prediction. This method uses Shapley~values~\cite{shapleyValuesOriginalPaper} from coalitional game theory. Each player is a feature or a coalition of features, and the game (payout) is the prediction. Shapley values are considered optimal because they satisfy the properties of efficiency, symmetry, dummy, and additivity. 

\subsubsection{Classifier Enhancing} The third approach consists of taking a well-established malware classifier for Android as a baseline and measuring its performance when enhanced with features related exclusively to cryptographic API. To this end, we chose \texttt{R-PackDroid}~\cite{maiorca_r-packdroid_2017}, an available learning-based classifier (trained on random forests) based on static features, and we expand its feature set by adding the cryptographic features described above. There are multiple reasons for which this system was chosen as a baseline: (i) It was initially designed to detect ransomware; (ii) It harvests a relatively small number of features; (iii) It features a high detection rate (the original paper documents over 97\% F1 score).  

Considering the characteristics described above, it would normally be challenging to improve the already strong performance of the system by adding more features. To keep some space for improvement when enhancing the classifier with cryptographic features, we decreased the number of \texttt{R-PackDroid} features from 211 to 10 in a controlled manner, leading to an F1 score of $76\%$. To measure the effect of cryptographic API features on the model, we replicated the following procedure $1000$ times: \emph{(i)} We sample 10 random features from \texttt{R-PackDroid}\footnote{The tuples were sampled in advance to avoid repetition.}; \emph{(ii)} We build a random forest classifier and measure its F1 score; \emph{(iii)} We enhance the 10 \texttt{R-PackDroid} features with all 189 cryptographic API features chosen by feature selection; \emph{(iv)} We use the expanded feature set to build another random forest classifier and measure its performance gain over the baseline classifier.

With the three strategies described above, we unveil the role of cryptographic API for malware detection, as will be shown in Section~\ref{sec:expresults:subsec:rq2}.

\section{\uppercase{Implementation notes}}
\label{app:implementation}

We now provide a more detailed description of each module, as well as the technical challenges that we had to face during the development of our system. Our tool is implemented in Python 3.8 and is provided as an open-source repository for further collaboration. Apart from reproducing our experiment, any other Android dataset can be plugged into our system to obtain a report about its cryptography usage. 

\subsection{Pre-processor}

The first task of the pre-processor is to obtain the decompiled Java source code of the input APKs. To get the object-oriented access to the source code, we instrumented the open-source tool \texttt{Androguard}~\cite{pythonsoftwarefoundationAndroguard}. As \texttt{Androguard} supports multiple decompilers, we made preliminary tests with various decompilers to verify that applications would be correctly parsed. The attained results showed that  \texttt{JADX} decompiler~\cite{jadx} is capable of the most mature recovery of the Java code, and hence was used in this study. For each APK, all \texttt{.java} classes in its \texttt{.dex} files were recovered. In some samples, a small fraction ($<1\%$) of classes did not survive the decompilation process. These classes were ignored from further processing. 

The second task of the pre-processor is to craft a list of third-party packages residing in the scrutinized APK. As mentioned in the previous section, this task is handled with the help of \texttt{LiteRadar} that was trained on a large dataset of Android applications, and can reliably fingerprint more than $29\,000$ third-party libraries. It should be stressed that the \texttt{LiteRadar} does not rely on package names, and can thus identify obfuscated packages as well. Using \texttt{LiteRadar}, we check every APK for the presence of third-party packages. Each package identified as third-party is then excluded from the cryptographic API analysis. 

\subsection{Crypto-extractor}

The crypto-extractor component executes two sub-tasks. The first objective is to gather a list of third-party \textit{cryptographic} libraries imported from the APK. For this scenario, we discriminate between \emph{(i)} Java cryptographic libraries and \emph{(ii)} native cryptographic libraries. In total, we searched for 23 distinct libraries. Their names, together with the reasons behind choosing them, can be found in Section~\ref{sec:expresults}.

The candidate list of native cryptographic libraries is then matched inside any of the following three import statements that load a native library directly from the source code: \texttt{ReLinker.loadLibrary}, \texttt{System.loadLibrary}, and \texttt{Native.loadLibrary} . The candidate list of Java cryptographic libraries is compared with the list of third-party packages identified earlier by \texttt{LiteRadar}. If any package appears in both lists, we note down the usage of the respective cryptographic library.

The second goal of the crypto-extractor is to collect comprehensive data about the cryptographic API usage. Although the packages \texttt{javax.crypto} and \texttt{java.security} contain more than 100 classes and interfaces, only some of those can reveal insights about the diversity of cryptography usage in the malware. We analyzed all classes in these packages and discarded out-of-scope instances to obtain \numanalyzedclasses classes for our analysis. Most of the diversity in the cryptographic API landscape can be explained by the study of \emph{object constructors}, and of their parameters. Whatever the developers' aim concerning cryptography is, they must first create a suitable object to address it. To give an example, when developers want to hash a file, they must first obtain the hash object by calling the constructor \texttt{MessageDigest.getInstance()}. When this constructor is called, e.g., with a string parameter \texttt{"SHA-256"}, this reveals the probable usage of SHA-256 hash function in the APK. We specified \numstatements constructors and parameters, and recorded all occurrences of these in the source code\footnote{While having the capability to capture such diverse landscape, in Section~\ref{sec:expresults} we present results only for 220 constructor variants from 8 classes, since the rest is used very rarely, and no conclusions can be drawn from such rare events.}. Specifically, we performed a line-by-line search of each of the user-defined classes. If the searched line contained any of the constructors, we note down its usage. By doing so, we collected a rough landscape of cryptography usage in the whole source code. This data was further refined and processed to draw the conclusions. 
 
Notably, constructors parameters can be obfuscated (and thus missed by our analysis - e.g., \texttt{MessageDigest.getInstance(a)}, where \texttt{a} is some variable). In this case, our system cannot properly parse such constructors. However, we argue that this limitation could only partially be statically solved, as even more advanced techniques (such as program slicing) can be easily defeated by more advanced obfuscation~\cite{maiorca15-cose,hoffmann16-sac}. Moreover, in large-scale scenarios, the problems introduced by the presence of some obfuscation are significantly outweighed by the dataset size. Nevertheless, for the sake of a fair analysis, we computed the exact fraction of obfuscated constructors for each of the cryptographic primitives we analyzed. 

\subsection{Evaluator and post-processing}

Once the overall JSON report for all APKs in the dataset is acquired, we post-process the results. Our tool automatically generates csv files of usage statistics and plots for all categories (and their sub-categories) described in Section~\ref{sec:methodology:subsec:formalization}. Apart from that, useful general information about the nature of the dataset is provided in the report. The resulting data serves as an input for the pipeline of a cryptographic learning model.

\subsection{System deployment}

The parallel processing of all 604 thousand samples took 16 days on 42 cores of Intel Xeon X7560, and each core consumed approximately $1.6\,$GB of RAM. That is an equivalent of processing 7 thousand APKs per 24 hours on a single CPU with 4 cores and 16\,GB of RAM, making the system well scalable. The subsequent post-processing of the JSON record to the form of the cryptography usage report takes 5 minutes on a regular laptop with 4 cores.

\section{\uppercase{Trends in cryptography}}
\label{sec:expresults}

\begin{figure*}
    \centering
    \begin{tikzpicture}
    \phantomsubcaption\label{fig:temporal:api_prevalence}
    \phantomsubcaption\label{fig:temporal:third_party_prevalence}
    \phantomsubcaption\label{fig:temporal:benign}
    \phantomsubcaption\label{fig:temporal:malicious}
        \begin{groupplot}[
        group style={
            group name=my plots,
            group size=4 by 1,
            xlabels at=edge bottom,
            ylabels at=edge left,
            horizontal sep=0.9cm,vertical sep=1.4cm,},
        xmin=2011.8,xmax=2020.2,
        xtick={2012, 2020},
        x tick label style={anchor=north, /pgf/number format/.cd, 1000 sep={}},
        yticklabel={$\pgfmathprintnumber{\tick}\%$},
        width=4.5cm,
        tick label style={font=\scriptsize},
        ]
        
        \nextgroupplot[xlabel={(a) cryptographic API}, ytick={10, 30, 50, 70}, legend style={nodes={scale=0.6, transform shape}, at={(0.01,0.01)}, anchor=south west}, cycle multi list=Dark2-5, ylabel={\% of APKs in dataset},]
        \addlegendentry{malicious}
        \addplot[mark=square*,thick, blue_shap, mark size=1.5pt] coordinates{(2012, 53.179) (2013, 66.348) (2014, 55.772) (2015, 48.577) (2016, 41.477) (2017, 23.780) (2018, 26.407) (2019, 16.536) (2020, 5.906)};
        
        \addlegendentry{benign}
        \addplot[mark=diamond*,thick, Dark2-A, mark size=2.5pt] coordinates {(2012, 27.378) (2013, 26.808) (2014, 30.681) (2015, 34.577) (2016, 37.519) (2017, 30.947) (2018, 45.746) (2019, 48.553) (2020, 49.627)};
        
        \nextgroupplot[xlabel={(b) third-party package}, ymax=102, ytick={0, 25, 50, 75, 100}, xtick={2012, 2020}, legend style={nodes={scale=0.6, transform shape}, at={(0.02,0.02)}, anchor=south west}]
        \addlegendentry{malicious}
        \addplot[mark=square*,thick, blue_shap, mark size=1.5pt] coordinates {(2012, 92.18) (2013, 91.81) (2014, 91.74) (2015, 80.90) (2016, 73.13) (2017, 85.52) (2018, 53.86) (2019, 18.05) (2020, 8.88)};
        \addlegendentry{benign}
        \addplot[mark=diamond*,thick, Dark2-A, mark size=2.5pt] coordinates {(2012, 79.09) (2013, 92.34) (2014, 96.88) (2015, 97.06) (2016, 98.71) (2017, 98.56) (2018, 98.50) (2019, 98.43) (2020, 97.15)};
        
        \nextgroupplot[xlabel={(c) Benign AES/DES}, ytick={0,5,10,15, 20}, legend style={nodes={scale=0.6, transform shape}, at={(0.02,0.98)}, anchor=north west}, xtick={2012, 2020}]
        \addlegendentry{AES}
        \addplot[mark=square*,thick, blue_shap, mark size=1.5pt] coordinates{(2012, 5.58) (2013, 4.97) (2014, 5.29) (2015, 9.80) (2016, 8.73) (2017, 8.61) (2018, 16.52) (2019, 15.38) (2020, 18.47)};
        \addlegendentry{DES}
        \addplot[mark=diamond*,thick, Dark2-A, mark size=2.5pt] coordinates {(2012, 1.71) (2013, 1.48) (2014, 1.07) (2015, 1.55) (2016, 1.80) (2017, 1.60) (2018, 2.37) (2019, 2.15) (2020, 0.91)};

        \nextgroupplot[xlabel={(d) Malicious AES/DES}, ytick={0,5,10,15}, legend style={nodes={scale=0.6, transform shape}}, xtick={2012, 2015, 2020}]
        \addlegendentry{AES}
        \addplot[mark=square*,thick, blue_shap, mark size=1.5pt] coordinates{(2012, 3.77) (2013, 6.19) (2014, 6.40) (2015, 11.02) (2016, 10.38) (2017, 3.86) (2018, 8.09) (2019, 2.22) (2020, 3.26)};
        \addlegendentry{DES}
        \addplot[mark=diamond*,thick, Dark2-A, mark size=2.5pt] coordinates {(2012, 11.14) (2013, 14.06) (2014, 10.87) (2015, 8.35) (2016, 5.82) (2017, 1.87) (2018, 6.20) (2019, 1.05) (2020, 0.60)};
        \draw [dashed] (axis cs:2015, 0) -- (axis cs:2015, 15);

        \end{groupplot}
    \end{tikzpicture}
    \caption{Time evolution of important dataset characteristics. The y-axis show the percentage of APKs exhibiting given feature, whereas the x-axis represents the time in years. In subplot (\subref{fig:temporal:api_prevalence}) we see the ratio of APKs in which we detected \textit{any} cryptographic API. Subplot (\subref{fig:temporal:third_party_prevalence}) shows APKs for which we detect any third-party package. These are the main artifacts of increasing obfuscation. Subplots (\subref{fig:temporal:benign}), (\subref{fig:temporal:malicious}) document the usage of AES vs. DES in benign, and malicious samples respectively. AES has been the most prevailing cipher suite in benign applications since 2012. On the contrary, DES was more popular in malware in previous years, only in 2015 being outran by AES.}
    \label{fig:temporal}
\end{figure*}
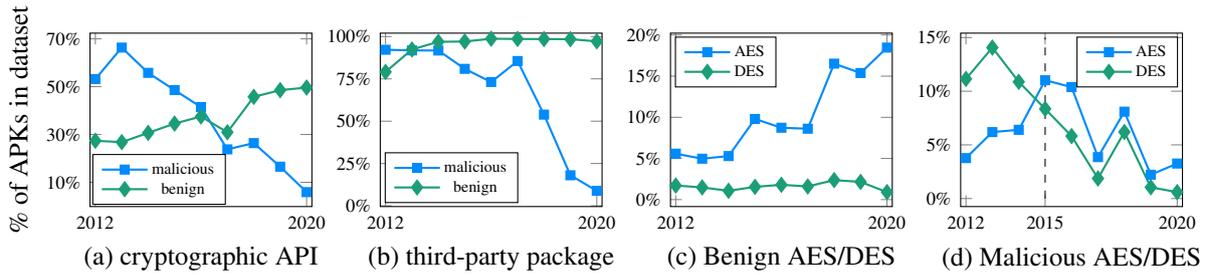

In this section, we answer RQ.1: \textit{Are there significant differences in how cryptography is employed in benign and malicious applications?} We report the most significant statistics we obtained with the API extraction methodology presented in Section~\ref{sec:methodology:subsec:pipeline}. First, we discuss the general prevalence of cryptographic API that can be extracted with static analysis by discussing the effects of obfuscation, the incidence of third-party packages, and the overall differences between benign and malicious applications. Then, we provide a more detailed focus on the distribution of cryptographic API in malicious applications. 

\subsection{General API Distribution}
\label{sec:expresults:subsec:rq1}

\subsubsection{Application Obfuscation}

The first interesting trend of this study is a decreasing ratio of malicious applications for which we detect usage of cryptographic API, as depicted in Figure~\ref{fig:temporal:api_prevalence}. However, such a ratio is not visible for benign applications. 
We conjecture that this drop does not represent a genuine decrease in usage of cryptographic functionality in time but rather a consequence of increasing ratio of obfuscated malicious applications. To confirm this hypothesis, we randomly sampled 4444 malicious applications from 2018-2020 that allegedly contain no cryptographic API or third-party libraries. We dissected them using commercial dynamic analysis tool \texttt{apklab}\footnote{Kindly provided by Avast, available at \url{http://apklab.io}.} and searched for clues of obfuscation. We identified that 98\% applications use some form of Android packer, with \texttt{jiagu} being most popular. Each packed application also uses reflection API and dynamic library loading, tools that prevent static analysis from registering cryptographic API call sites. We also report that 83\% of applications use some form of encryption API (AES being the most prevalent, followed by RSA), often to decrypt application resources. Such reduced prevalence of crypto API constitutes a limitation of our study, further discussed in Section~\ref{sec:discussion}.

\subsubsection{Third-party Packages and Crypto Libraries}
Another closely related trend is the dropping ratio of third-party packages captured by \texttt{LiteRadar} in malicious applications. Before 2018, we documented a high ratio of malware employing third-party packages ($86\%$). Starting with 2018, this ratio quickly drops as depicted in Figure~\ref{fig:temporal:third_party_prevalence}.
Similar to obfuscation, this drop is not evident in benign applications. Overall, \texttt{LiteRadar} was able to identify at least one third-party package in $94.6$\% of analyzed goodware with little variance between years (see Figure~\ref{fig:temporal:third_party_prevalence}). On average, $8$ packages were identified in benign APKs. This underpins the importance of robust third-party package detection. In contrast to prior work that did not consider third-party package filtering, we discarded over $4$ million third-party packages with at least $44$ thousand unique package names from the analysis. 

\begin{table}[ht]
    \centering
    \begin{tabular}{|c |}
        \hline
        3rd-party cryptographic libraries \\
        \hline
        Java \\
        \hline
        \texttt{whispersystems/curve25519} \\
        \texttt{guardianproject/netcipher} \\
        \texttt{springframework/security/crypto} \\
        \texttt{gnu/crypto} \\
        \texttt{apache/shiro/crypto} \\
        \texttt{rsa/crypto} \\
        \texttt{keyczar} \\
        \texttt{jasypt} \\
        \texttt{googlecode/gwt/crypto} \\
        \texttt{sqlcipher} \\
        \texttt{spongycastle} \\
        \texttt{bouncycastle} \\
        \texttt{facebook/crypt} \\
        \hline
        native \\
        \hline
        \texttt{crypto-algorithms} \\
        \texttt{libgcrypt} \\
        \texttt{monocypher} \\
        \texttt{PolarSSL} \\
        \texttt{tint-AES-C} \\
        \texttt{xxHash} \\
        \texttt{libsodium} \\
        \texttt{openssl} \\
        \texttt{libressl} \\
        \texttt{wolfssl} \\
        \hline
      \end{tabular}
      \vspace{4mm}
    \caption{List of 23 third-party cryptographic libraries that were searched in each of the studied samples. The Java libraries were identified using the \texttt{LiteRadar} tool, whereas the native libraries were matched as case-insensitive regular expressions inside the import statements \texttt{ReLinker.loadLibrary}, \texttt{System.loadLibrary}, \texttt{Native.loadLibrary} from the decompiled source code.}
    \label{tab:third_party_libs}
\end{table}

Apart from Android API, cryptographic functions can also be delivered by third-party libraries, typically adopted to integrate functionality missing in system-based libraries. 
To our best knowledge, no curated list of third-party cryptographic libraries for Android exists. We manually selected \numjavalibs Java and \numnativelibs native candidate libraries to be searched for. These candidates were found through Google search engine and in popular databases~\cite{bauerAndroidArsenal}, and their fit was confirmed by manual inspection. Although this process is inherently incomplete and some libraries could have been missed out, we argue that this is not a practical limitation since the most prevalent libraries were unlikely to be missed, and even these are rarely used.  The full list of third-party cryptographic libraries searched in the samples can be found in Table~\ref{tab:third_party_libs}. 

We identified only 796 benign and 198 malicious applications that import third-party cryptographic libraries. Of the studied libraries, \texttt{sqlcipher} was most popular in goodware (found in $622$ samples), and \texttt{keyczar} was most popular in malware (found in $124$ samples). The ratio of these applications has been stable throughout the studied timeline. As for the native libraries, not a single call to a native cryptographic library was detected in the malicious dataset, and merely 91 imports of \texttt{OpenSSL} occurred in the benign dataset. 

From these results, it is possible to observe that third-party cryptographic libraries are not widely used in Android applications. This aspect demonstrates that attackers often resort to standard crypto functionalities provided by system libraries (that can use various backends, e.g., BouncyCastle). 

\subsubsection{Crypto API in Goodware and Malware}~\label{sec:expresults:comparison}

We now describe the general prevalence of cryptographic API in the dataset presented in Section~\ref{sec:methodology:subsec:dataset} by showing the differences between malicious and benign applications and comparing our results with two studies conducted on benign datasets. For this comparison, we employed: \textit{(i)} A dataset collected in 2012 as a part of the study CryptoLint~\cite{egeleEmpiricalStudyCryptographic2013} that we refer to as CryptoLint-B12; \textit{(ii)} A dataset collected in 2016 as a part of Binsight study~\cite{muslukhovSourceAttributionCryptographic2018} that we refer to as Binsight-B16. To avoid temporal data drift, we cast four subsets of our dataset: Androzoo-B12, Androzoo-M12, Androzoo-B16, and Androzoo-M16, limited to malicious (M), and benign (B) samples from years 2012 (12), and 2016 (16). As explained in Section~\ref{sec:methodology}, our goal is to analyze only cryptographic APIs contained in user-defined code. From this respect, both \cite{egeleEmpiricalStudyCryptographic2013,muslukhovSourceAttributionCryptographic2018} employ weaker methodologies to filter third-party libraries, relying on whitelisting and package names. Conversely, our approach of \texttt{LiteRadar} filtering captures the code written by the application authors more reliably. The numbers drawn from the Androzoo datasets serve as conservative estimates, with the real number of cryptographic API call sites even higher. The overall comparison with benign datasets is depicted in Table~\ref{tab:benign_malicious}. It can be seen that the malicious datasets have a dramatically higher density of cryptographic API call sites than their benign counterparts. 

\begin{table*}
    \centering
    \begin{tabular}{| c | c | c | c |  }
        \hline
        Dataset &  \#APKs  & \#User-def. call sites & \#User-def. call sites/10k samples \\
        \hline
        CryptoLint-B12  & $145\,095$  & $20\,967$   & $1445$        \\
        BinSight-B16    & $115\,683$  & $78\,163$   & $7006$        \\
        Androzoo-B12    & $39\,838$   & $81\,698$   & $20\,507$     \\    
        Androzoo-B16    & $37\,493$   & $124\,705$  & $33\,260$     \\ 
        Androzoo-M12    & $39\,767$   & $125\,225$   & $31\,489$    \\
        Androzoo-M16    & $39\,325$   & $208\,625$  & $53\,051$     \\
        \hline
      \end{tabular}
      \vspace{2mm}
    \caption{Comparison of cryptographic API spread in benign vs. malicious datasets. The last column normalizes by the size of the datasets, allowing for direct comparison.}
    \label{tab:benign_malicious}
\end{table*}

\begin{table*}
    \centering
    \begin{tabular}{|c | c | c | c | c | c | c | }
        \hline
        Dataset & AES & DES & 3DES & RC4 & Blowfish & Unknown \\[1mm]
        \hline
        CryptoLint-B12  & \textbf{58.9\%} & $19.0$\% & $8.8$\% & $0.4$\% & $1.9$\% & $10.9$\% \\
        BinSight-B16    & \textbf{64.4\%} & $14.3$\% & $1.1$\% & $2.1$\% & $0.9$\% & $17.2$\% \\
        Androzoo-B12    & \textbf{52.4\%} & $16.9$\% & $3.8$\% & $0$\% & $0.0$\% & $26.8$\% \\
        Androzoo-B16    & $\mathbf{59.0\%}$ & $12.2$\% & $2.0$\% & $0.1$\% & $0.0$\% & $26.8$\% \\
        Androzoo-M12    & $12.1$\% & \textbf{56.0\%} & $0.9$\% & $0.0$\% & $0.0$\% & $31.0$\% \\
        Androzoo-M16    & \textbf{45.1\%} & $22.8$\% & $2.1$\% & $0.0$\% & $0.0$\% & $30.0$\% \\
        \hline
      \end{tabular}
      \vspace{2mm}
    \caption{Distribution of symmetric ciphers in benign and malicious datasets with AES dominating all but Androzoo-M12.}
    \label{tab:m16_b16}
\end{table*}

\textbf{CryptoLint-Anndrozoo Comparison.} The CryptoLint-B12 dataset resulted from scanning $145\,095$ samples for the presence of cryptographic API (and its misuse).
The study concluded that $15\,134$ ($10.4$\%) of APKs contain some cryptographic call sites. However, the subsequent BinSight study attributed $79.5$\% of these call sites to the ignored third-party packages, showing that the original CryptoLint study suffered from overcounting. 

In contrast, we report that $27.4\%$ of Androzoo-B12 contains cryptographic API call sites and nearly twice as much malware from Androzoo-M12 ($53.1$\%), too. This highlights the extensive use of cryptographic API in malicious applications compared to the benign landscape. A closer examination of symmetric ciphers in Table~\ref{tab:m16_b16} also reveals considerable differences between malicious and benign datasets. AES dominates benign datasets with $58.9$\% in CryptoLint-B12 and $52.4\%$ in Andozoo-B12. The situation is strikingly different in the malicious dataset. In Androzoo-M12, the most popular primitive is DES with $56$\% of call sites, followed by AES ($12.1$\%) and 3DES ($0.9$\%). We provide more in-depth comparison of individual ciphers and their modes of operation in Appendix~\ref{app:symmetric_encryption}.

\textbf{BinSight-Androzoo Comparison.}
The BinSight paper aimed to answer what proportion of cryptographic API misuse can be attributed to third-party packages. The authors identified $638$ distinct third-party packages in $115\,683$ unique samples in BinSight-B16, relying on the package name as an identifier. The authors attributed at least $90.7$\% of the call sites to third-party packages, underlying the need for their robust detection. Even after we discarded $9\,870$ third-party packages from Androzoo-M16, the malicious dataset still contains much more cryptographic API in the user-authored codebase. Again, the relations between Androzoo-M16 and BinSight-B16 are depicted in Table~\ref{tab:benign_malicious}. Interestingly, in 2016, AES was dominant in Androzoo-M16 as well with $45.1\%$ of call sites, followed by DES ($22.8\%$). We depict the time evolution of AES vs. DES in Androzoo dataset in Figure~\ref{fig:temporal}, showing that it was only in 2015 when AES outran DES in malicious applications.

\subsection{Crypto API Categories in Malware}
\label{sec:expresults:subsec:malw}

Apart from the comparison to benign applications, we also report a broad view of the distribution of cryptographic API in \emph{malicious} applications, concentrating on the years 2012-2018, for which we dispose of representative samples not clouded with high ratio of obfuscated applications.

Table~\ref{tab:categories} illustrates that the majority of call sites from this period can be attributed to hash functions ($66$\%) and symmetric encryption ($26$\%), which leaves the rest of the categories rarely used. Nevertheless, we comment on our findings in all categories, observing the time evolution trends and showing the most prevailing primitives. We could not attribute $21\%$ of the identified constructors to the exact cryptographic primitive (partial obfuscation) during our experiments. We still manage to pinpoint their presence and even their category, as the system-based API calls are challenging to obfuscate entirely.

\textbf{Hash Functions.}
The hash functions are by far the most popular category of cryptographic API in malicious applications, as they are present in 40\% of all studied APKs and responsible for $424\,858$ call sites in our dataset. Interestingly, the majority of the call sites resort to primitives MD5 or SHA-1 that were already shown to be broken~\cite{wang2005break,stevens2017first}. 
Specifically, MD5 can be attributed to more than $80$\% of these call sites and does not lose any of its popularity in time. This may suggest that MD5 is either not meant to provide secure integrity protection for the authors or that the developers are unaware of its weakness. The time evolution of SHA-1 and SHA-256 points to the former case. Indeed, the overall dominant SHA-1 (almost $16$\% of call sites) is gradually decreasing over time in favor of the more secure SHA-256 ($3$\% overall). In 2018, SHA-256 was present in more APKs (708) than SHA-1 (528). This phenomenon 
can mean that, when secure integrity protection is needed, more secure SHA-256 is nowadays being selected instead of SHA-1. Still, MD5 is preferred by the malware creators for other use cases. Apart from the hash functions mentioned above, only SHA-512 and SHA-384 are represented in the dataset, but these are responsible for less than $1000$ call sites in total. 

\textbf{Symmetric Encryption.}
A large portion of the symmetric encryption API landscape was already described in Section~\ref{sec:expresults:comparison}, but some important aspects were yet omitted. 
Overall, our dataset contains $165\,994$ symmetric encryption call sites distributed in approximately $20$\% of APKs. A large portion of the call sites ($26$\%) is obfuscated.
Besides AES and DES, only 3DES is used in more than 1000 APKs. We also report that the concept of password-based encryption is applied merely in 837 APKs. A closer look at the encryption modes offers an interesting perspective. Our observations confirm that the authors favor the default constructors (\texttt{"AES"} and \texttt{"DES"}) compared to constructors that specify encryption mode and padding (e.g., \texttt{"AES/CBC/PKCS5PADDING"}).
The default constructors fall back into the ECB mode with PKCS\#7 padding, which is (under most circumstances) considered insecure~\cite{menezes1996handbook}.

\textbf{Public-key Encryption.}
The only asymmetric encryption scheme appearing in the Android API is RSA. The RSA encryption occurs in approximately $1.55$\% of all APKs in our dataset. Until 2013, RSA appeared very rarely, but then it peaked within two years to almost $1\,800$ APKs in 2015. 

\begin{table}
    \centering
    \begin{tabular}{|l | c | c  | c| }
        \hline
        Category            & \#call sites  & \%obfusc. & \%APK \\[1mm]
        \hline
        Hash functions      & $424\,858$    & $16.8$\%        & $39.7$\% \\
        Symmetric enc.      & $165\,994$    & $25.9$\%       & $19.4$\% \\ 
        Public-key enc.     & $13\,262$     & $25.9$\%       & $1.5$\%\\ 
        Digital sig. alg.  & $17\,505$     & $81.4$\%       & $4.5$\% \\
        MAC                 & $11\,661$     & $46.4$\%       & $3.0$\% \\
        PRNGs               & $10\,381$     & $6.6$\%        & $2.9$\% \\
        Key agreement       & $87$          & $29.9$\%       & $0$\% \\
        \hline
        Sum                 & $646\,018$    & $21.5$\%       & $44.6$\% \\
        \hline
      \end{tabular}
      \vspace{2mm}
    \caption{Popularity distribution of cryptographic API categories. The ratio of APKs for symmetric encryption and RSA is approximate, since one cannot differentiate  the obfuscated constructors of these two categories. Note that approximately 1\% of APKs contain cryptographic API outside of these categories.}
    \label{tab:categories}
\end{table}

\textbf{Digital Signature Algorithms.}
Surprisingly, digital signature algorithms occupy $4.5$\% of the malicious APKs and are present in $17\,505$ call sites. Considering possible applications of digital signature primitives in malware, this constitutes a rather large number. Despite the highest obfuscation rate among categories ($81.43$\%), the \texttt{SHA1withRSA} primitive is responsible for almost $80\%$ of the unobfuscated call sites. As in the case of the hash functions, \texttt{SHA256withRSA} is on the rise in time, first appearing in 2015 and steadily increasing the fraction of APKs it appears in ever since. Still, in 2018 it is less than four times probable to appear compared to \texttt{SHA1withRSA}. 

While multiple schemes supporting elliptic curves over RSA or DSA are offered in the API, these are explicitly specified only in 19 APKs in total, with the first use appearing in 2014.

\textbf{MAC Algorithms.}
The situation with MAC algorithms is similar to that of digital signatures algorithms. The MAC systems are responsible for $11\,661$ call sites and are present in $3\%$ of APKs. Still,  a large portion ($46.4$\%) of the call sites are obfuscated. Nevertheless, only two functions are called in more than 1\% of the call sites -- \texttt{HMACSHA1} and \texttt{HMACSHA256}. The former is heavily dominant throughout the studied timeline, being responsible for $70\%$ of the MAC call sites.

\textbf{PRNGs.}
The functionality of PRNGs is utilized in nearly 3\% of the APKs, being responsible for $10\,381$ call sites. A relatively small fraction of the call sites ($6.6$\%) are obfuscated, and virtually all unobfuscated call sites (over $90$\%) can be attributed to \texttt{SHA1PRNG}. 

\textbf{Key Agreement Protocols.}
The key agreement API's functionality consists purely of the Diffie-Hellman protocol (DH) for key exchange. Concerning the DH protocol parameters, we can only differentiate between the use of DH over finite fields or elliptic curves. The key agreement API appears only in 53 APKs over the nine years, occupying 87 call sites in total. 36 of the APKs use elliptic curves, whereas 20 APKs use obfuscated calls. Interestingly, only a single APK was detected to be explicitly using DH over finite fields. The dominant use of elliptic curves over finite fields is in contrast with the situation in digital signature algorithms. 

Worth noting, we did not thoroughly explore how cryptographic primitives were employed in the \emph{context} of the applications (e.g., to send SMS, encrypt data, et cetera). This analysis is extremely complex due to the variety of application contexts, and it is hardly feasible with static analysis. However, to give readers possible directions about the motivations for using cryptography in malware, we manually inspected a small subset of samples during our study. For the categories defined in Section~\ref{sec:methodology:subsec:formalization} we documented the following use-cases: \emph{(i)} \textit{Hash functions} are generally used to fingerprint the attributes of a device (IMEI, Android version, etc.), to hash whole file or string, or to construct home-brew MACs or signature primitives; \emph{(ii)} \textit{Public-key encryption} was witnessed to provide hybrid encryption, or to construct digital signature algorithms from its basic blocks; \emph{(iii)} \textit{Symmetric encryption} is used to encrypt files, as well as strings, and to obfuscate expressions directly in the source code. We also witnessed the use of \textit{PRNG} to generate random keys (often with static seeds) or to provide nonces for more complex scenarios; \emph{(iv)} Both \textit{key agreement protocols} and \textit{digital signature algorithms} were found to empower more complex network protocols, e.g., SASL (Simple Authentication and Security Layer~\cite{sasl}); \emph{(v)} We report no surprising use-cases for \textit{MAC} primitives that serve their original purpose of data authentication.

\section{\uppercase{Machine learning and cryptographic API}}
\label{sec:expresults:subsec:rq2}

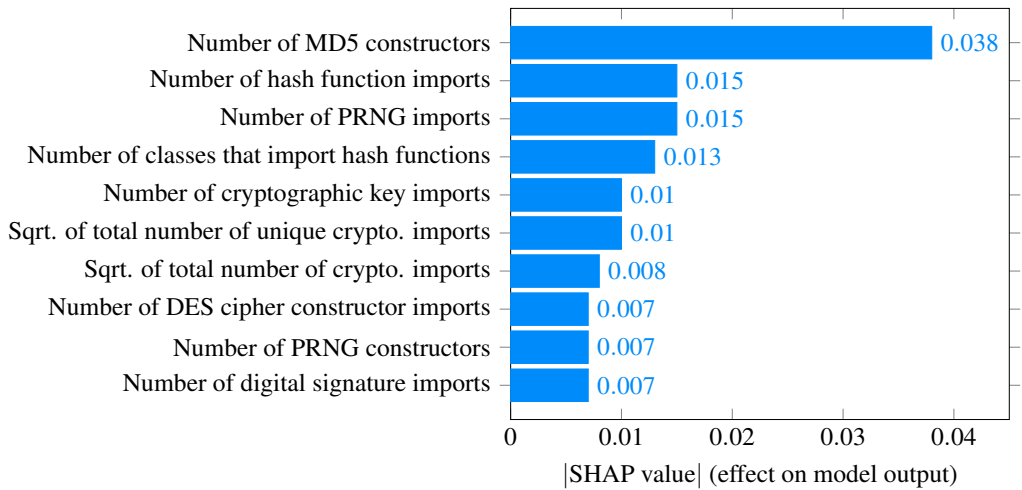
\begin{figure*}
    \centering
    \begin{adjustbox}{width=0.85\textwidth}
        \begin{tikzpicture}
            \begin{axis}[ 
            xbar, xmin=0,
            bar width=0.45cm,
            xlabel={$|\textrm{SHAP value}|$ (effect on model output)},
            symbolic y coords={
                {Number of digital signature imports},
                {Number of PRNG constructors},
                {Number of DES cipher constructor imports},
                {Sqrt. of total number of crypto. imports},
                {Sqrt. of total number of unique crypto. imports},
                {Number of cryptographic key imports},
                {Number of classes that import hash functions},
                {Number of PRNG imports},
                {Number of hash function imports},
                {Number of MD5 constructors}},
            ytick=data,
            nodes near coords, 
            nodes near coords align={horizontal},
            nodes near coords style={
                    /pgf/number format/fixed,
                    /pgf/number format/precision=5
                    },
            xticklabel style={
                    /pgf/number format/fixed,
                    /pgf/number format/precision=5
            },
            scaled x ticks=false,
            xmax = 0.045,
            ]
            \addplot[blue_shap,fill=blue_shap] coordinates {
                (0.038,{Number of MD5 constructors}) 
                (0.015,{Number of hash function imports}) 
                (0.015,{Number of PRNG imports})
                (0.013,{Number of classes that import hash functions})
                (0.010,{Number of cryptographic key imports})
                (0.010,{Sqrt. of total number of unique crypto. imports})
                (0.008,{Sqrt. of total number of crypto. imports})
                (0.007,{Number of DES cipher constructor imports})
                (0.007,{Number of PRNG constructors})
                (0.007,{Number of digital signature imports})};
            \end{axis}
        \end{tikzpicture}
    \end{adjustbox}
    \caption{A representation of 10 most influential features (their $\left | \textrm{SHAP} \right |$ values are high, averaged over all samples). These represent the spots in cryptographic API with the largest difference in usage between malicious and benign samples. The x-axis shows the average effect on the model output in either direction. The model outputs values from 0 (benign) to 1 (malicious).}
    \label{fig:global_feature_importance}
\end{figure*}

In this section, we answer RQ.2: \textit{Can information about cryptography improve Android malware detection?} To do so, we analyze the outcome of the experiments outlined in Section~\ref{sec:methodology:subsec:motivation}.

\subsubsection{Cryptography-Based Learning Model}
We trained a random forest model based only on cryptography-related features described in Section~\ref{sec:methodology:subsec:motivation} and compared its performance to \texttt{R-PackDroid}. To obtain a valid comparison, we replicated the experimental setup of the original \texttt{R-PackDroid} paper~\cite{maiorca_r-packdroid_2017}, taking 10 thousand applications divided 50:50 into benign/malicious, and split 50:50 into training/test set. 
Our classifier achieved $62.4\%$ F1 score on the malicious samples (see also Table~\ref{tab:rpackdroid_experiment}), showing that cryptographic information is discriminant enough to separate malicious from benign samples. Even though that \texttt{R-PackDroid} performs significantly better than our system\footnote{Remember that our goal was not to build a better classifier but to show that it is possible to distinguish between malicious and benign Android applications only by their cryptographic API usage.}, our classifier was able to correctly identify 88/180 malicious samples that were misclassified as benign by \texttt{R-PackDroid} (with all 211 features). This shows that cryptographic API can assist the classification of samples that would otherwise fly under the radar of existing classifiers. 

\subsubsection{Explanations of Decisions}
In Figure~\ref{fig:global_feature_importance}, we present the 10 most influential features of the cryptographic-API classifier, based on the SHAP values calculated for the whole dataset (global explanation) according to the methodology from Section~\ref{sec:methodology:subsec:motivation}. Note that the model outputs values from 0 (benign) to 1 (malicious), and the expected value of the model on a balanced dataset is hence $0.5$. The SHAP value of a feature thus represents a deviation from this expected value after inspecting a particular feature. 

It is rather interesting to see that the usage of certain hashing functions is discriminative w.r.t. maliciousness of the samples. More specifically, weak hash functions (MD5) are especially used in malicious samples (as also reported by the analysis in Section~\ref{sec:expresults:subsec:rq1}), and they constitute an important indicator of maliciousness. Additionally, the classifier observes the general number of imported cryptographic functions. An increasing number of imported functions lead to an increasing suspicion of maliciousness.

\begin{figure*}
    \centering
    \includegraphics[width=1\textwidth]{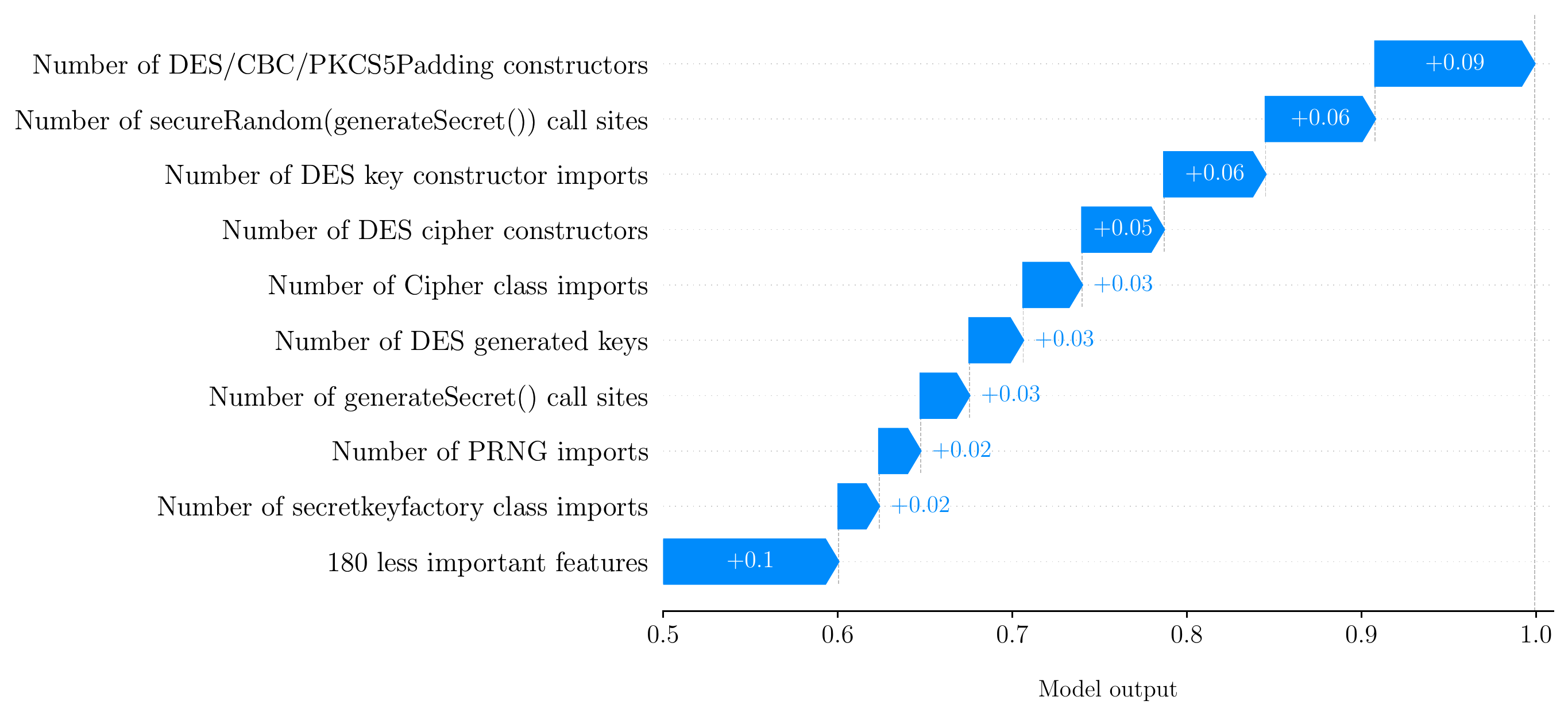}
    \caption{Local impact of 10 most influential features w.r.t the models' prediction \textit{on a particular sample}. The shown APK is a malware of SMSpay type. It exfiltrated DES-encrypted data through SMS. Apart from the features, the figure also depicts how the value of each feature shifts the models' output from a neutral score $0.5$ to the final output $1$ that labels the APK as malware.}
    \label{fig:local_feature_importance}
\end{figure*}

Concerning the local explanation, we present an example related to the malware with MD5 hash \texttt{e1001da40929df64443f6d4037aa3a9f}. VirusTotal classifies this sample as a riskware of type SMSpay. By extracting the local SHAP values (Figure~\ref{fig:local_feature_importance}), it is possible to see significant importance of a DES encryption that steers the classifiers' decision towards maliciousness. Driven by this explanation, we manually disassembled the sample and looked for the usage of DES-related cryptographic API. We found that, in this case, DES is used to encrypt sensitive information, such as the phone device id, which is then subsequently exfiltrated to a remote server. This detail is especially useful to attract the attention of the analyst towards malicious operations carried out by the sample. Also, note that this sample employs name obfuscation, and the required effort to carry out similar analysis without such guidance would be higher.  

\begin{table}
    \centering
    \begin{tabular}{| c | c | c | }
        \hline
        Classifier  & \# features & F1 score\\
        \hline
        Cryptographic API       & 189       & $62.40\%$ \\
        RPackDroid (full)       & 211       & $92.71\%$ \\
        RPackDroid (limited)    & 10        & $76.47\%$ \\
        RPackDroid + cryptoAPI  & $10+189$  &  $81.33\%$ \\
        \hline
      \end{tabular}
      \vspace{2mm}
    \caption{Comparison of performance of malware classifiers without and with cryptographic API features. The performance metrics measure the result of the malicious samples. Enhancing the limited RPackDroid with cryptographic features causes $4.18\%$ increase of precision and $5.61\%$ increase of recall on the Androzoo dataset, projecting into $4.86$\% F1 score increase.}
    \label{tab:rpackdroid_experiment}
\end{table}

\subsubsection{Enhancing Existing Classifier with Cryptographic API Features}
\label{sec:expresults:subsec:rq2:subsubsec:classifier}

In the original work, \texttt{R-PackDroid} achieved 97\% F1 score distinguishing between three classes (ransomware in addition to malware/benign). Leveraging the source code of authors, we achieved $92.71\%$ F1 score with \texttt{R-PackDroid} on the Androzoo dataset. Thus, enhancing this classifier would not bring a big performance gain. Hence, we decreased the number of features employed by the classifier to 10, reducing F1 score to $74.47\%$ when averaged over different 10-tuples of features. We then added the cryptographic features following the methodology from Section~\ref{sec:methodology:subsec:motivation}.

We document an increase of classifiers' performance when enhanced with crypto features. Our results demonstrate that, while the margin for improvement is not huge, adding cryptographic features significantly improved both recall ($+5.61\%$), and precision ($+4.18\%$) of the classifier, which in turn increases the F1 score by $+4.86\%$. See the summary in Table~\ref{tab:rpackdroid_experiment}.

\section{\uppercase{Discussion and limitations}}
\label{sec:discussion}

Our main goal was to provide a comprehensive overview of the role of cryptography in the analysis of Android applications and malware detection. As explained in Section~\ref{sec:expresults:subsec:rq1}, we found a significant amount of obfuscation in recent malicious applications, which prevented us from providing detailed information about their usage of cryptography. We recognize that this is an inherent limitation of static analysis. While other static techniques such as program slicing may provide additional insights, we believe that dynamic analysis is the only reliable way to defeat dynamic code loading and other types of packing. However, it is infeasible to analyze more than half million of APKs with dynamic analysis. Hence, we believe that our proposed approach constitutes the right balance between effectiveness, precision, and analysis time.

We also point out that possible biases can be present in the data we analyzed. In particular, we did not have control over the contents of the Androzoo dataset. According to the indications provided by the Androzoo authors~\cite{allixAndroZooCollectingMillions2016}, we can safely rule out the presence of possible duplicates for applications coming from the same sources (e.g., the same stores). We point out that the risk of finding duplicates across stores is significantly lower for malicious applications than benign ones. Nevertheless, even if such duplicates were found, their number should not influence a large-scale analysis.

\section{\uppercase{Related work}}
\label{sec:relwork}

Most of the research on cryptographic API in Android is mainly focused on benign applications where the ultimate goal is to mitigate its misuse. Several steps are needed to achieve this, and the respective works usually treat one or two steps at a time. We can summarize these steps as follows:
\emph{(i)} Inferring the rules of cryptographic API misuse;
\emph{(ii)} Evaluation of cryptographic API misuse; 
\emph{(ii)} Attribution of cryptographic API misuse;
\emph{(iv)} Automatic cryptographic API repairs. 
The following paragraphs discuss the related research for all these steps mentioned above.

\textbf{Inferring Rules of Cryptographic API Misuse.}
In the area of inferring the rules of cryptographic API misuse, the goal is to create a list of specifications for developers and researchers that imply the insecure use of cryptography. Such rules can be crafted manually as done in~\cite{egeleEmpiricalStudyCryptographic2013,chatzikonstantinouEvaluationCryptographyUsage2016,shuaiModellingAnalysisAutodetection2014}. However, this approach does not scale well, leading to the works~\cite{paletovInferringCryptoAPI2018,gaoNegativeResultsMining2019} that attempt to infer these rules from git commits, conjecturing that newly introduced commits typically eliminate security vulnerabilities from the code. Surprisingly, Paletov et al.~\cite{paletovInferringCryptoAPI2018} reported success with this approach, whereas chronologically later work~\cite{gaoNegativeResultsMining2019} commends against the initial assumption. 

\textbf{Evaluation of Cryptographic API Misuse.}
After having a set of rules that suggest security violations at hand, it is vital to explore these violations in the Android applications market. While more powerful dynamic analysis is employed in~\cite{chatzikonstantinouEvaluationCryptographyUsage2016,shuaiModellingAnalysisAutodetection2014} to show that more than half of the examined applications violate the static set of rules, the application dataset is relatively small (size $<100$). On the contrary, the static analysis approach used by Egele et al. in~\cite{egeleEmpiricalStudyCryptographic2013} allowed examining a large dataset of $145$ thousand benign applications to reveal that $10.4$\% of them uses some form of cryptography. $88\%$ of such applications were found to violate some rule of secure cryptography usage. These results were confirmed by a later study~\cite{muslukhovSourceAttributionCryptographic2018} that gathered a new dataset of $109$ thousand APKs that contain at least one cryptographic API call and showed the analogical proportion of insecure applications.
Static rules were substituted by a more sophisticated definition language in~\cite{krugerCrySLExtensibleApproach2018} where 10\,000 Android applications were analyzed and misuses detected in over $95$\% of cases. 

Some of the solutions above are impractical to run against large projects due to many false positives. This is treated by \texttt{CryptoGuard}~\cite{rahaman_cryptoguard_2019} that prunes the alerts to achieve $98\%$ precision and is successfully run against real-world projects. As of early 2022, an open-source \texttt{CRYLOGGER}~\cite{piccolboni_crylogger_2021} can well complement \texttt{CryptoGuard}, as it is based on dynamic analysis and was tested on a sufficiently large dataset (approx. 1800 applications). In 2021, the first systematic evaluation study~\cite{ami_why_2021} was published that allows to measure the quality of such detectors and reveals many flaws in their design or implementation.

Concentrating on the TLS protocol, the work~\cite{fahlWhyEveMallory2012} from 2012 analyzed $13$ thousand Android applications to reveal inadequate TLS usage in $8$\% of cases. The authors also managed to launch $41$ MiTM attacks against selected applications. Iterating on this effort, an article~\cite{why_eve_still_loves_android} studied Network Security Configuration files\footnote{\url{developer.android.com/training/articles/security-config}} in Android. The authors revealed that $88\%$ of applications employing custom settings downgrade the security compared to the default configuration. Also, the authors penetrated Google Play safeguards that are supposed to protect from publishing applications vulnerable to MiTM.

\textbf{Attribution of Cryptographic API Misuse.}
\label{sec:related_work:attribution}
Reliable third-party package detection is central for attribution of misuse. This problem has been addressed, e.g., in~\cite{wangWuKongScalableAccurate2015,maLibRadarFastAccurate2016,libscout,libscout} where matching algorithms were proposed to reliably detect third-party libraries. As already discussed, a systematic review~\cite{zhan_automated_2020} then compared these detectors from various perspectives confirming that \texttt{LibRadar} is superior to others when used for large-scale analysis due to result quality comparable with the most precise tools, yet running much faster.

\textbf{Automatic Cryptographic API Repairs.}
More distant to our research are papers that concentrated on automatic cryptographic API misuse repairs. From this area of research, we refer the reader to~\cite{maCDRepAutomaticRepair2016,zhangEmbroideryPatchingVulnerable2017}.

\textbf{Study of Cryptography in Android Malware.}
We point out that all the aforementioned research on the Android platform did not concentrate on cryptographic API in malicious applications or its comparison to the benign landscape. Our work in this paper aims to fill this gap in knowledge. 

\section{\uppercase{Conclusions and future work}}
\label{sec:conclusions}

In this paper, we provide a longitudinal analysis of how cryptographic API was typically employed in Android malware in the years 2012-2020. We analyzed \numsamplesdetailed applications and extracted over 1 million call sites that have been extensively analyzed by extracting various statistics. The attained results showed the following:
\begin{enumerate}
    \item \textit{Use of weak hash functions}. Most malicious applications featuring cryptographic routines resorted to weak MD5 hash functions.
    \item \textit{Late transition from DES to AES}. It is worth pointing out one aspect of comparison with benign samples. In the symmetric cipher category, malware authors adopted weak DES to modern AES only in 2015, while AES was the most popular cipher in benign samples already in 2012.
    \item \textit{Very limited use of third-party cryptographic libraries}. Our analysis showed that Android application authors favor using system-based libraries to deliver cryptographic functionality.
    \item \textit{Contrast between malicious and benign usage of cryptography}.
    Our study shows that cryptographic API is generally more frequent in malware than in benign samples (in relative measures).
\end{enumerate}
Building a cryptography-based machine learning model, we showed a significant difference in cryptography deployment between benign and malware samples. We demonstrated that a learning-based model based on cryptography alone could separate between benign and malicious samples with good performance. Moreover, we showed that explaining the decision of the classifier can expose intriguing links between cryptography and malicious actions that are typically carried out by malware. In particular, these techniques constitute a valid resource to guide the analysts towards discovering critical characteristics of the examined malicious samples. Moreover, cryptographic features until now neglected can be employed to further improve state-of-the-art malware detectors.

Our results open door to various follow-up work. For instance, it would be interesting to cluster malware samples into families based on its usage of cryptography. Likewise, future work can more closely focus on understanding \emph{for what purpose} specific crypto-routines are employed in Android malware, thus better understanding and profiling the characteristics of malware authors. This could, for instance, confirm or dismiss our conjecture that weak hash functions are not used for integrity-critical operations in malware. Last but not least, one could employ more powerful dynamic analysis to check if our findings hold also for packed and obfuscated applications.

\section*{\uppercase{Acknowledgements}}
This work was partially supported by the project PON AIM Research and Innovation 2014–2020 - Attraction and International Mobility, funded by the Italian Ministry of Education, University and Research; and by the European cybersecurity pilot CyberSec4Europe. Vashek Matyas was supported by Czech Science Foundation project 
GA20-03426S. Adam Janovsky was supported by Invasys company. We are grateful to Jonas Konecny who ran the initial machine learning experiments. We also thank Avast for providing the dynamic-analysis tool \url{apklab.io}.

\bibliographystyle{apalike}
{\small
\bibliography{references}}

\appendix

\begin{table*}[ht]
    \centering
    \begin{tabular}{|c | c | c | }
        \hline
        Symmetric encryption scheme & Androzoo-M12 & CryptoLint-B12 \\[1mm]
        \hline
        DES* & 6356 & 741 \\
        DES/CBC/PKCS5Padding & 1203 & 205 \\
        AES* & 924 & 4803 \\
        AES/CBC/PKCS5Padding & 786 & 5878 \\
        DESede/ECB/PKCS5Padding & 231 & 473 \\
        AES/ECB/PKCS5Padding & 122 & 443 \\
        DESede* & 107 & 501 \\
        DES/ECB/PKCS5Padding & 93 & 221 \\
        DES/ECB/NoPadding & 68 & 1151 \\
        AES/CBC/NoPadding & 43 & 468 \\
        AES/ECB/NoPadding & 41 & 220 \\
        AES/CBC/PKCS7Padding & 37 & 235 \\
        AES/CFB8/NoPadding & 24 & 104 \\
        AES/ECB/PKCS7Padding & 1 & 155 \\
        \hline
        Sum AES where freq. $>100$ & 1832 & 12306 \\
        Sum DES where freq. $>100$ & 7559 & 2318 \\
        Sum DESede where freq. $>100$ & 338 & 974 \\
        Sum where freq. $>100$ & 9729 & 15598 \\
        \hline
      \end{tabular}
      \vspace{4mm}
    \caption{Comparison of distribution of symmetric encryption schemes in malicious vs. benign applications (Androzoo-M12 and CryptoLint-B12). The frequency of malicious encryption schemes was normalized to fit the size of the benign dataset. In the benign set, only the schemes with frequency $>100$ were taken. There is no prevalent malicious scheme (freq. $>100$) that would not appear in the benign dataset. The default schemes marked with * symbol fall back into the ECB mode with PKCS7 padding.}
    \label{tab:b12_m12}
\end{table*}

\section{Detailed comparison of symmetric ciphers between Androzoo-M12 and CryptoLint-B12}
\label{app:symmetric_encryption}

Table~\ref{tab:b12_m12} displays an in-depth comparison between symmetric encryption schemes in the datasets CryptoLint-B12 and Androzoo-M12. It should be stressed that even though the absolute number of call sites in CryptoLint-B12 is higher ($15\, 598$)  than in Androzoo-M12 ($9729$), this comparison is severely skewed by the overall distribution characteristics of CryptoLint-B12 vs. Androzoo-M12. In other words, it takes $145$ thousand of benign applications (where only each fifth call originates from user-defined codebase) to get $15$ thousand calls, whereas $34$ thousand of malicious applications would provide a similar number of symmetric encryption API call sites.

\end{document}